\RequirePackage{fixltx2e}
\documentclass[reprint,superscriptaddress,amsmath,amssymb,aps,prl,floatfix]{revtex4-1}

\usepackage{color}
\usepackage{graphicx}
\usepackage{bm}
\usepackage{hyperref}
\usepackage[mathlines]{lineno} 
\usepackage{microtype}
\usepackage{times}
\usepackage[varg]{txfonts}
\usepackage{booktabs}
\usepackage{hyperref}
\usepackage[nolist,nohyperlinks]{acronym}
\usepackage{pdfpages}

\definecolor{cblue}{RGB}{55,126,184}
\hypersetup{colorlinks=true,linkcolor=cblue,citecolor=cblue,urlcolor=cblue}

\newcommand{\subref}[2]{\ref{#1}\hyperref[#1]{#2}}
\newcommand{\shortsection}[1]{\emph{#1:}}

\renewcommand{\vec}[1]{\bm{\mathbf{#1}}}

\newcommand{\vhat}[1]{\vec{\hat{#1}}}

\newcommand{\hc}{\textrm{h.c.}}

\newcommand{\bra}[1]{\left \langle #1\right|}
\newcommand{\ket}[1]{\left| #1  \right \rangle}

\newcommand{\avg}[1]{\langle #1 \rangle}

\newcommand{\tss}[1]{\textsubscript{#1}}
\newcommand{\tsp}[1]{\textsuperscript{#1}}

\newcommand{\kb}{k_{\rm B}}

\newcommand{\meV}{\ \textrm{meV}}
\newcommand{\K}{\ \textrm{K}}
\newcommand{\mK}{\ \textrm{mK}}

\newcommand{\baybzno}{Ba\tss{3}Yb\tss{2}Zn\tss{5}O\tss{11}}
\newcommand{\abo}[2]{{#1}\tss{2}{#2}\tss{2}O\tss{7}}
\newcommand{\yb}{Yb\tsp{3+}}

\newcommand{\Ei}{E_{\rm i}}

\newacro{BYZO}[BYZO]{\baybzno{}}
\newacro{DM}[DM]{{Dzyaloshinskii-Moriya}}
\newacro{BP}[BP]{breathing pyrochlore}
\newacro{INS}[INS]{inelastic neutron scattering}
\newacro{AIAO}[AIAO]{all-in/all-out}

\begin{document}

\title{Anisotropic exchange within decoupled tetrahedra in the quantum breathing pyrochlore \baybzno{}}
\thanks{This manuscript has been authored by UT-Battelle, LLC under Contract No. DE-AC05-00OR22725 with the U.S. Department of Energy.  The United States Government retains and the publisher, by accepting the article for publication, acknowledges that the United States Government retains a non-exclusive, paid-up, irrevocable, world-wide license to publish or reproduce the published form of this manuscript, or allow others to do so, for United States Government purposes.  The Department of Energy will provide public access to these results of federally sponsored research in accordance with the DOE Public Access Plan (http://energy.gov/downloads/doe-public-access-plan).}

\author{J. G. Rau}
\email{jeff.rau@uwaterloo.ca}
\affiliation{Department of Physics and Astronomy, University of Waterloo, Ontario, N2L 3G1, Canada}

\author{L. S. Wu}
\email{wul1@ornl.gov}
\affiliation{Quantum Condensed Matter Division, Oak Ridge National Laboratory, Oak Ridge, TN-37831, USA}

\author{A. F. May}
\affiliation{Materials Science \& Technology Division, Oak Ridge National Laboratory, Oak Ridge, TN-37831, USA}

\author{L. Poudel}
\affiliation{Quantum Condensed Matter Division, Oak Ridge National Laboratory, Oak Ridge, TN-37831, USA}
\affiliation{Department of Physics \& Astronomy, University of Tennessee, Knoxville, TN-37966, USA}

\author{B. Winn}
\affiliation{Quantum Condensed Matter Division, Oak Ridge National Laboratory, Oak Ridge, TN-37831, USA}

\author{V. O. Garlea}
\affiliation{Quantum Condensed Matter Division, Oak Ridge National Laboratory, Oak Ridge, TN-37831, USA}

\author{A. Huq}
\affiliation{Chemical \& Engineering Materials Division, Oak Ridge National Laboratory, Oak Ridge, TN 37831, USA}

\author{P. Whitfield}
\affiliation{Chemical \& Engineering Materials Division, Oak Ridge National Laboratory, Oak Ridge, TN 37831, USA}

\author{A. E. Taylor}
\affiliation{Quantum Condensed Matter Division, Oak Ridge National Laboratory, Oak Ridge, TN-37831, USA}

\author{M. D. Lumsden}
\affiliation{Quantum Condensed Matter Division, Oak Ridge National Laboratory, Oak Ridge, TN-37831, USA}

\author{M. J. P. Gingras}
\affiliation{Department of Physics and Astronomy, University of Waterloo, Ontario, N2L 3G1, Canada}
\affiliation{Perimeter Institute for Theoretical Physics, Waterloo, Ontario, N2L 2Y5, Canada}
\affiliation{Canadian Institute for Advanced Research, 180 Dundas Street West, Suite 1400, Toronto, ON, M5G 1Z8, Canada}

\author{A. D. Christianson}
\affiliation{Quantum Condensed Matter Division, Oak Ridge National Laboratory, Oak Ridge, TN-37831, USA}
\affiliation{Department of Physics \& Astronomy, University of Tennessee, Knoxville, TN-37966, USA}

\date{\today}

\begin{abstract}
The low energy spin excitation spectrum of the breathing pyrochlore \baybzno{} has been investigated with inelastic neutron scattering.  Several nearly resolution limited modes with no observable dispersion are observed at $250 \mK$ while, at elevated temperatures, transitions between excited levels become visible.  To gain deeper insight, a theoretical model of isolated \yb{} tetrahedra parametrized by four anisotropic exchange constants is constructed.  The model reproduces the inelastic neutron scattering data, specific heat, and magnetic susceptibility with high fidelity. The fitted exchange parameters reveal a Heisenberg antiferromagnet with a very large Dzyaloshinskii-Moriya interaction. Using this model, we predict the appearance of an unusual octupolar paramagnet at low temperatures and speculate on the development of inter-tetrahedron correlations.
\end{abstract}

\maketitle

Frustrated or competing interactions have been repeatedly found to be at the root of many unusual phenomena in condensed matter physics~\cite{springer-2011-frustrated, kivelson-tarjus-2005-glass, watanabe-2011-nuclear-pasta, balents-2010-spin-liquids, diep-2013-frustrated-spin-systems}. By destabilizing conventional long-range order down to low temperature, frustration in magnetic systems can lead to many exotic phases; from unconventional multipolar~\cite{santini-2009-multipolar-review, starykh-2015-unusual-review} and valence bond solid orders~\cite{springer-2011-frustrated, balents-2010-spin-liquids} to disordered phases such as classical and quantum spin liquids~\cite{springer-2011-frustrated,balents-2010-spin-liquids}. Significant attention has been devoted to understanding \emph{geometric} frustration where it is the connectivity of the lattice that hinders the formation of order. Recently, however, magnets frustrated not by geometry but by competing interactions have become prominent for the novel behaviors that they host.  Such competing interactions might be additional isotropic exchange acting beyond nearest neighbors~\cite{iqbal-thomale-2015-paramagnetism,  wills-2012-kapellasite, bombardi-carretta-2004-square}, biquadratic or other multipolar interactions~\cite{mila-penc-2006-biquadratic}. One possibility attracting ever increasing interest is that competing strongly \emph{anisotropic} interactions may stabilize a wide range of unusual phenomena.

\begin{figure}[htp]
    \centering
    \includegraphics[width=0.75\columnwidth]
    {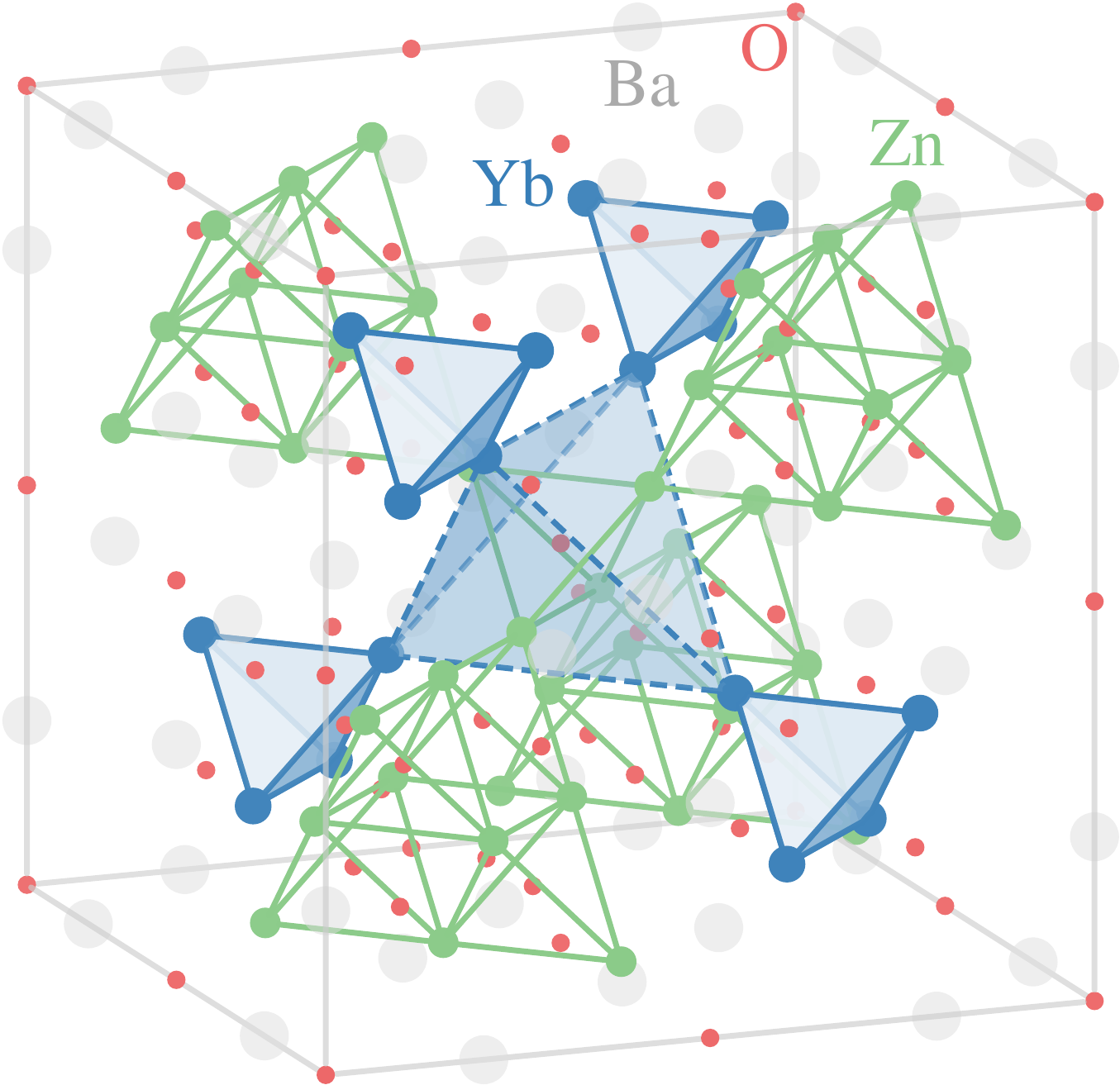}
    \caption{\label{fig:unitcell}
    Crystal structure of \baybzno{} (cubic space group $F\bar{4}3m$, no. 216). Each \yb{} ion is part of a large and small tetrahedron in the breathing pyrochlore lattice.}
\end{figure}

An exciting research direction in the latter context concerns itself with so-called ``quantum spin ice"~\cite{gingras-mcclarty-2014-quantum-spin-ice}. This quantum spin liquid can be stabilized by perturbing classical spin ice with additional anisotropic transverse exchange interactions that induce quantum fluctuations. Particularly interesting is the potential realization of such physics in the rare-earth pyrochlores \abo{R}{M}~\cite{molavian-gingras-2007-quantum, tanaka-onoda-2010-quadrupole, savary-balents-2011-quantum-ice}, where R is a trivalent $4f$ rare-earth ion, and M is a non-magnetic tetravalent transition metal ion, such as  M=Ti, Sn or Zr. These materials can be described in terms of pseudo spin-$1/2$ degrees of freedom interacting via anisotropic exchanges~\cite{gingras-mcclarty-2014-quantum-spin-ice, savary-balents-2011-quantum-ice}, where the effective spin-$1/2$ maps the states of the crystal-electric field ground doublet of the rare-earth ion. These materials display a wealth of interesting phenomena, from the possibility of quantum~\cite{savary-ross-2012-order, zhitomirsky-moessner-2012-order, wong-gingras-2013-quantum} order-by-disorder physics in \abo{Er}{Ti}~\cite{rau-gingras-2015-virtual}, unconventional ordered states~\cite{chang-onoda-2012-higgs,stewart-wills-2004-partial}  as well as several candidates for quantum spin liquids~\cite{gardner-1999-cooperative-frustrated,  kimura-2013-quantum-fluctuations}. In many of these compounds, the physics is very delicate, showing strong sample to sample variations~\cite{ross-2012-stuffed-pyrochlore} or sensitivity to very small amounts of disorder~\cite{taniguchi-2013-spin-liquid,kadowaki-2015-quadrupole}. Consequently, an accurate determination of the effective model is crucial in making definite progress in this area. This is particularly true in cases where the idealized disorder-free material may find itself in the vicinity of a transition between competing semi-classical ground states~\cite{wong-gingras-2013-quantum, robert-petit-2015-dynamics, jaubert-gingras-2015-multi}

Given the critical importance played by the precise value of the anisotropic exchanges, a number of experiments have been aimed at determining those couplings~\cite{savary-balents-2011-quantum-ice, savary-ross-2012-order}. There is, unfortunately, much difficulty in obtaining accurate values for these couplings stemming from two key limitations. First, only approximate methods are available to relate the model to experiment, restricting comparisons to regimes where the theory becomes controlled, such as in high magnetic field~\cite{savary-balents-2011-quantum-ice, savary-ross-2012-order,hayre-gingras-2013-quantum} or at high-temperature~\cite{thompson-gingras-2011-anisotropic, applegate-gingras-2012-quantum, hayre-gingras-2013-quantum,oitmaa-singh-2013-order}. Second, to avoid over-fitting the experimental data, one must work with a reasonable number of fitting parameters; for example restricting to a subset of the allowed interactions by ignoring interactions beyond nearest neighbors or possible multi-spin interactions~\cite{rau-gingras-2015-virtual}. Even in \abo{Yb}{Ti}, where the latter concern is largely absent, there currently remains no consensus on the values of the anisotropic exchange parameters~\cite{savary-balents-2011-quantum-ice, robert-petit-2015-dynamics}. At the present time, a reference rare-earth pyrochlore-like compound with solely bilinear anisotropic interactions and for which essentially exact methods can be employed to compare with experimental data, is badly needed to cement the validity of the effective spin-$1/2$ description of such materials.

In this Letter, we study \ac{BYZO}, a so-called \ac{BP} compound~\cite{kimura-2014-breathing, haku-2015-breathing}, which provides an ideal platform for understanding such anisotropic exchange models.  As shown in Fig.~\ref{fig:unitcell}, BYZO consists of small tetrahedra with a short nearest-neighbor bond distance $r_<\sim 3 \AA$ connected by large tetrahedra with size $r_>\sim 6 \AA$. Because of the large ratio $r_>/r_< \sim 2$, the inter-tetrahedron couplings are expected to be small compared to the intra-tetrahedron couplings, leading to effectively decoupled small tetrahedra. This can be compared to the Cr-based \ac{BP} compounds, where the small and large tetrahedra only differ in size by $\sim 5\%$~\cite{okamoto-2013-breathing, tanaka-2014-breathing-pyrochlore, okamoto-2015-breathing}. To characterize BYZO spectroscopically, we have investigated its low energy spin excitations using \ac{INS}. We confirm the picture of nearly independent tetrahedra, seeing nearly resolution limited dispersion-less modes at low temperatures.  This \ac{INS} data, combined with the thermodynamic measurements of Ref.~[\onlinecite{kimura-2014-breathing}], allows for a complete and unambiguous determination of the the effective model for BYZO. We find that a single tetrahedron pseudo-spin model can quantitatively account for all of the current experimental data on BYZO, determining the four anisotropic exchanges as well as the $g$-tensor. In addition to the antiferromagnetic Heisenberg exchange postulated in Ref.~[\onlinecite{kimura-2014-breathing}], we find that significant \ac{DM} exchange is needed to obtain the correct level structure determined from \ac{INS}. The fitted exchange parameters are far from the spin ice limit recently considered in Ref.~[\onlinecite{savary-kee-kim-chen-2015-breathing}] or the purely Heisenberg limits studied in Ref.~[\onlinecite{benton-shannon-2015-breathing}]. Instead, we find the ground state of each tetrahedron is doubly degenerate, consistent with the residual entropy observed experimentally at $T\sim 300$ mK~\cite{kimura-2014-breathing}. These $E$-doublets are nearly non-magnetic, carrying both a scalar spin-chirality as well as octupolar, all-in/all-out moments. The state of \ac{BYZO} at currently studied base temperatures is thus a type of ``octupolar paramagnet" without significant inter-tetrahedron correlations. Notwithstanding the broad agenda of accurately determining the anisotropic exchanges in rare-earth pyrochlore materials, the complete characterization of the single-tetrahedron model should provide a useful guide for further experimental studies of BYZO and other \ac{BP}s. Specifically, we estimate that the inter-tetrahedron correlations could begin to set in below $500\mK$, at the edge of currently explored temperatures, possibly leading to interesting new physics~\cite{tsunetsugu-2001-quantum-pyrochlore, tsunetsugu-2001-order-pyrochlore, kotov-zhitomirsky-2004-weak-dimer, kotov-2004-patterns,kotov-elhajal-2005-spin-liquid} in this material.

\begin{figure*}[htp]
    \centering
    \includegraphics[width=\textwidth]
    {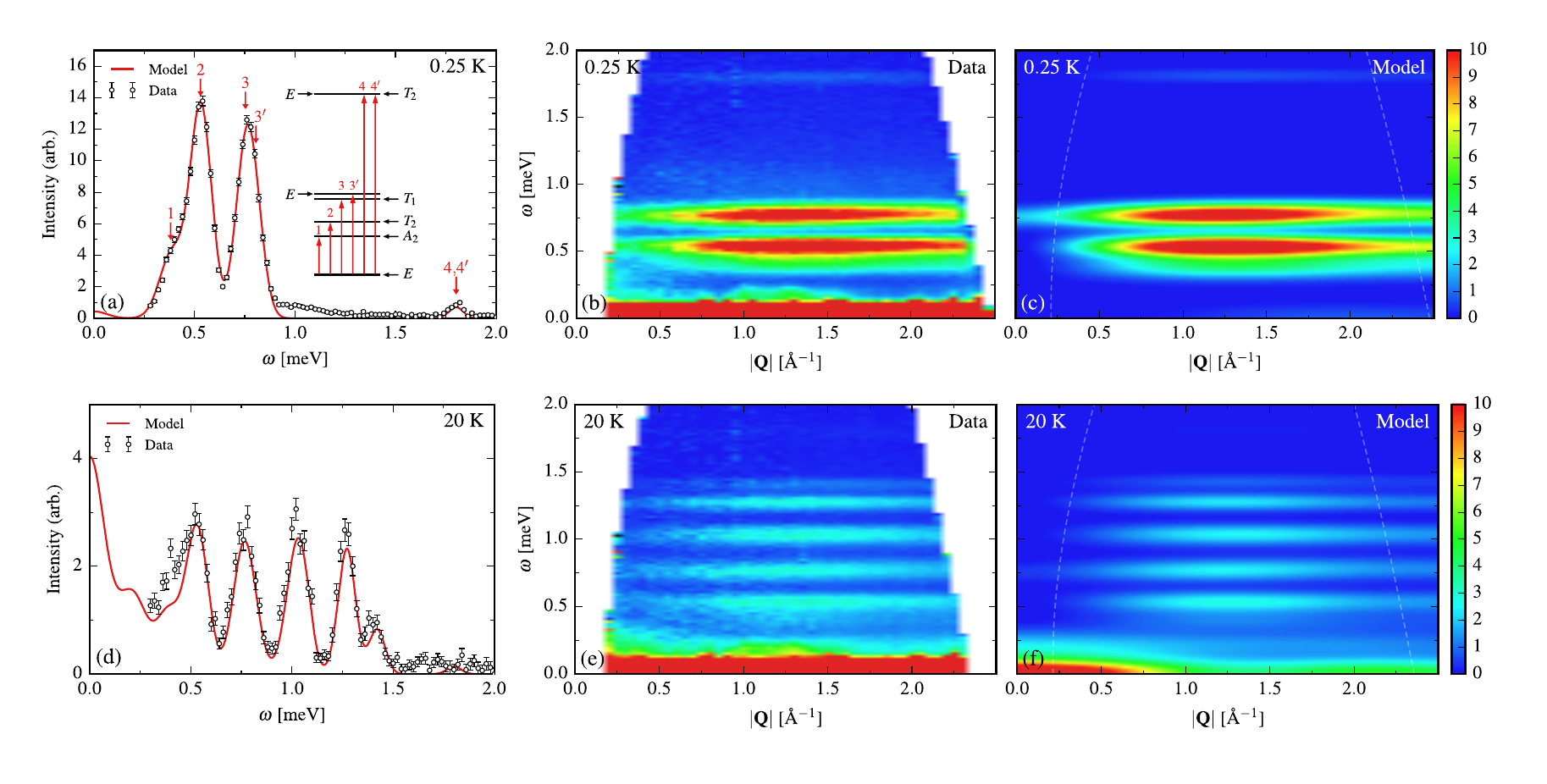}
    \caption{\label{fig:neutron}
    \ac{INS} data ($\Ei =3.8 \meV$) and comparison to our theoretical model at
    (a-c) $0.25 \K$ and (d-f) $20 \K$. The overall theoretical intensity scale was fit using the constant wavevector cut (a) at $0.25\K$. A Gaussian broadening with energy dependence following the experimental energy resolution function~\cite{supp} was included in the theoretical calculation.
    (a,d) Cut of the \ac{INS} data averaged over the window $1.25 \AA^{-1} < |\vec{Q}| < 1.35 \AA^{-1}$ at (a) $0.25\K$  and (b) $20\K$. Results for model of Eq.~(\ref{eq:model}) with fitted parameters of Eq.~(\ref{eq:best}) are also shown. (Inset,a) An illustration of the level structure of the single-tetrahedron model and the positions of the transitions from the ground doublet into the excited states.
    (b,e) Intensity map of the powder averaged \ac{INS} data at (c) $0.25\K$ and (d) $20\K$. The excitations are nearly dispersion free over the full $|\vec{Q}|$ range. (c,f) Model calculations for the parameters of Eq.~(\ref{eq:best}) are shown at (e) $0.25\K$ and (f) $20\K$. The \yb{} form factor was evaluated in the dipole approximation, as given in Ref.~[\onlinecite{inttab}].}
\end{figure*}

\shortsection{Experimental results} Polycrystalline samples of \ac{BYZO} were synthesized by solid-state reaction in Al$_2$O$_3$ crucibles~\cite{supp}. The resulting samples were characterized by specific heat and magnetization measurements~\cite{supp}.  The structure was studied via neutron powder diffraction utilizing the POWGEN~\cite{powgen2011} diffractometer at the Spallation Neutron Source at Oak Ridge National Laboratory~\cite{supp}.  These measurements confirm the previously reported cubic structure~\cite{struct1,struct2,kimura-2014-breathing} (space group $F\bar{4}3m$, no. 216) with lattice parameter $a$=13.47117(3) at 10 K and $a$=13.48997(3) at 300 K.

To explore the low energy spectrum of magnetic excitations in \ac{BYZO}, \ac{INS} data was collected using the HYSPEC spectrometer~\cite{hyspec} at the Spallation Neutron Source at Oak Ridge National Laboratory. Measurements were performed at 0.25, 10, and 20 K utilizing a $^3$He refrigerator, with fixed incident neutron energies of $\Ei$ = 3.8 \meV{}, 7.5 \meV{}  and 15 \meV{}.

\ac{INS} measurements with $\Ei$ = 3.8 meV at $0.25$ and $20 \K$ are shown in Fig.~\ref{fig:neutron}. The data at $0.25 \K$ (Fig.~\subref{fig:neutron}{(a)} and \hyperref[fig:neutron]{(b)}) exhibits several well-defined modes with no observable dispersion. The $|\vec{Q}|$-dependence of the inelastic scattering intensity exhibits a broad peak centered near $|\vec{Q}| = 1.3 \AA^{-1}$ (see Fig.~\subref{fig:neutron}{(b)} and the Supplemental Material~\cite{supp}). The width in energy of the modes is close to instrumental resolution~\cite{supp}. At elevated temperatures (Fig.~\subref{fig:neutron}{(b)} and \hyperref[fig:neutron]{(d)}), three new excitations become visible resulting from transitions between excited states.

The origin of the observed low energy excitations appears to be modes originating from \emph{decoupled} Yb tetrahedra. Several pieces of evidence support this assertion. Low lying crystal field levels can be excluded as the origin of these modes as three higher energy crystal field levels are experimentally observed (the maximum  number for \yb{}) with the lowest lying level at $\sim 38 \meV$~\cite{haku-2015-breathing, supp}.  The magnetic susceptibility and specific heat data do not show any signs of long range magnetic order down to $0.38 \K$~\cite{kimura-2014-breathing,supp} that would indicate correlations between the small tetrahedra.  Examination of the elastic scattering at $0.25 \K$ is consistent with this conclusion, revealing no indication of long range magnetic order. Finally, the lack of dispersion suggests that these modes arise primarily from isolated tetrahedra and that the interactions connecting the tetrahedra are weak. We note that there is a weak and broad feature at $\sim 1 \meV$. We have been unable to identify the origin of this feature, but note that it has a $|\vec{Q}|$-dependence~\cite{supp} distinct from that of the other nearly resolution limited modes.

\shortsection{Theoretical model} We now use these experimental observations, along with the thermodynamic data from Ref.~[\onlinecite{kimura-2014-breathing}] to construct a model of \ac{BYZO}. Given the dispersion-less modes seen in the INS, and the large ratio $r_>/r_< \sim 2$ between the large and small tetrahedron sizes, we expect isolated ${\rm Yb}_4$ tetrahedra to provide a very good description of the low energy physics. Each of the four \yb{} ions has a Hund's rule ground state of ${}^2 F_{7/2}$, with the $J=7/2$ manifold strongly split by the $C_{3v}$ ($3m$) crystalline electric field environment. Since this energy scale is very large, $\sim 38 \meV$~\cite{haku-2015-breathing}, relative to the expected scale of the intra-tetrahedron interactions, only the ground doublet is relevant at low temperatures. The two states of this doublet define an effective pseudo-spin $\vec{S}_i$ at each of the four \yb{} sites. This pseudo-spin is related to the magnetic moment $\vec{\mu}_i$ at each site through the $g$-factors, $g_z$ and $g_{\pm}$, present due to the local $C_{3v}$ symmetry. Explicitly,
\begin{equation}
\label{eq:moment}
    \vec{\mu}_i \equiv  \mu_B \left[ g_{\pm} \left(\vhat{x}_i S^x_i +  \vhat{y}_i S^y_i\right) + g_z \vhat{z}_i S^z_i\right],
\end{equation}
where $(\vhat{x}_i, \vhat{y}_i, \vhat{z}_i)$ are the local axes of tetrahedron site $i$ ~\cite{supp}. Regardless of the detailed composition of the ground doublet, since $J=7/2$, the interactions between the \yb{} are expected to be anisotropic and, a priori, not necessarily near the Ising or the Heisenberg limit~\cite{rau-2015-magnitude}. Symmetry strongly constrains their form; each \yb{}-\yb{} bond has symmetry $C_{2v}$ ($2mm$) and each small ${\rm Yb}_4$ tetrahedron has full tetrahedral symmetry $T_d$ ($\bar{4}3m$)~\cite{struct2,kimura-2014-breathing}. Assuming an effective spin-1/2 doublet \footnote{The case of a dipolar-octupolar doublet ($\Gamma_5 \oplus \Gamma_6$) is in principle possible as well, with a different anisotropic exchange model ~\cite{chen-hermele-2014-octupolar}. We find such a model does not provide a good description of the specific heat or magnetic susceptibility of \ac{BYZO} and thus consider only an effective spin-1/2 ($\Gamma_4$) doublet. This is consistent with the $\Gamma_4$ ground doublet found in Ref.~[\onlinecite{haku-2015-breathing}] by fitting the observed crystal field excitations.}, there are therefore four allowed anisotropic exchange interactions~\cite{savary-balents-2011-quantum-ice}, taking the form
\begin{align}
\label{eq:model}
    & H_{\rm eff} \equiv \sum_{i=1}^4 \sum_{j<i} \left[
        J_{zz} S^z_i S^z_j - J_{\pm}\left(S^+_i S^-_j+S^-_i S^+_j\right)+\right.  \nonumber \\
        & J_{\pm\pm} \left(\gamma_{ij} S^+_i S^+_j+\hc \right)+
        \left. J_{z\pm}  \left(
            \zeta_{ij} \left[ S^z_i S^+_j+ S^+_i S^z_j \right]+ \hc
            \right)
    \right],
\end{align}
where the bond dependent phases $\gamma_{ij}$ and $\zeta_{ij}$ are defined in the Supplemental Material~\cite{supp}. The spectrum of this Hamiltonian is partly determined by tetrahedral symmetry. The four-pseudo-spin states break into the irreducible representations $A_2 \oplus 3E  \oplus T_{1} \oplus 2T_{2}$ under the action of the tetrahedral group. This gives a level structure of a singlet ($A_2$), three doublets ($E$) and three triplets ($T_1$ or $T_2$).  From the observed residual entropy~\cite{kimura-2014-breathing}, it seems plausible that the ground state of the tetrahedron is an $E$ doublet, which gives an entropy of $\kb \ln(2)/4 \sim 0.1733 \kb$ / \yb{}.

\shortsection{Best fit parameters} The model of Eq.~(\ref{eq:model}), supplemented with the definition of the moment in Eq.~(\ref{eq:moment}), is determined by the six parameters $J_{zz}$, $J_{\pm}$, $J_{\pm\pm}$, $J_{z\pm}$, $g_z$ and $g_{\pm}$. To fix these parameters, we perform a fit to the specific heat and susceptibility data of Ref.~[\onlinecite{kimura-2014-breathing}] and a cut of the \ac{INS} data averaged over the range $1.25 \AA^{-1} < |\vec{Q}| < 1.35 \AA^{-1}$ at $0.25\K$.  This is a global fit, minimizing squared differences between experimental and theoretical values from each set of experimental data simultaneously. For the specific heat, we fit only the data below $5\K$ to minimize the influence of the subtraction of the lattice contribution, while the susceptibility data up to $30\K$ is used. \footnote{Fitting only the specific heat and susceptibility from Ref.~[\onlinecite{kimura-2014-breathing}] does not produce a unique fit, but many equally good fits. However, these differ significantly when including constraints that arise from fitting the \ac{INS} data}.  Three additional fitting parameters were included; a constant shift of the susceptibility, $\chi_0$, to account for the Van Vleck and diamagnetic core contributions of the \yb{} ions, the intensity scale of the \ac{INS} cut and the overall scale of the Gaussian broadening used in the theoretical \ac{INS} intensity~\cite{supp}. Further details of the fitting methodology and comparisons to experimental data can be found in the Supplemental Material~\cite{supp}.

\begin{figure}[tp]
    \centering
    \includegraphics[width=\columnwidth]
    {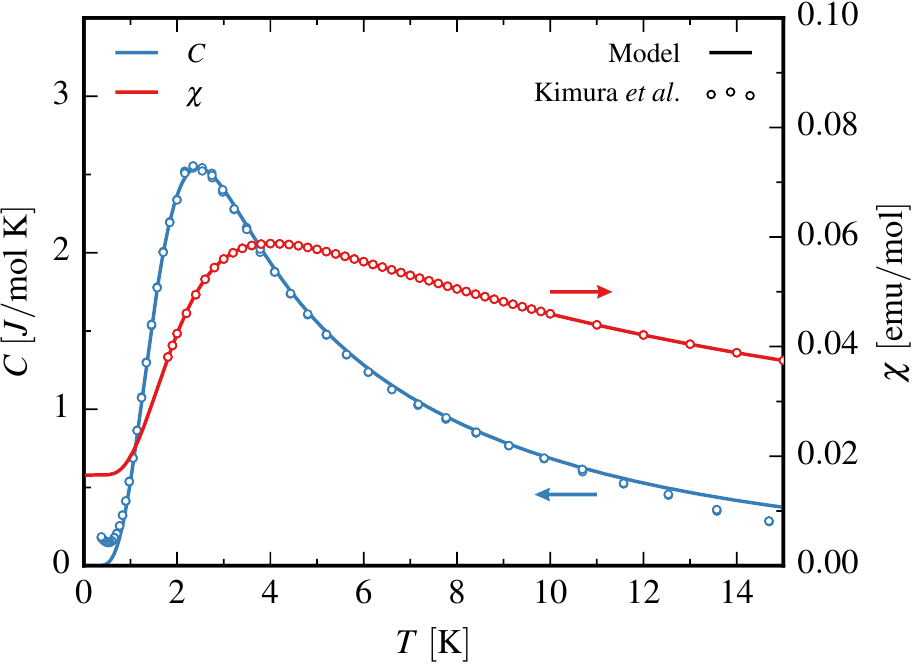}
    \caption{
    Comparison of the (magnetic) specific heat, $C$, and susceptibility, $\chi$, of~\citet{kimura-2014-breathing} to the model of Eq.~(\ref{eq:model}) with fitted parameters of Eq.~(\ref{eq:best}). A constant shift $\chi_0$ was included in the fit of the susceptibility to account for the Van Vleck and diamagnetic core contributions of the \yb{} ions.}
    \label{fig:therm}
\end{figure}

From this analysis we find a unique best fit which provides excellent agreement with \emph{all} of the known experimental data on \ac{BYZO}. The best fit parameters are
\begin{align}
\label{eq:best}
  J_{zz}     &= -0.037 \meV, &
  J_{\pm}    &= +0.141 \meV, \nonumber\\
  J_{\pm\pm} &= +0.158 \meV, &
  J_{z\pm}   &= +0.298 \meV, \nonumber\\
  g_{\pm}    &=  2.36, &
  g_{z}      &=  3.07.
\end{align}
Comparison to the specific heat and susceptibility is shown in Fig.~\ref{fig:therm}. Agreement with both is excellent; small differences can be seen in the specific heat at higher temperatures, likely due to some uncertainty in the subtraction of the lattice contribution.  Comparison to a cut of the \ac{INS} data at $0.25\K$ is shown in Fig.~\subref{fig:neutron}{(a)}, along with an illustration of the level structure of the single tetrahedron model with the parameters of Eq.~(\ref{eq:best}). The level structure matches very well with the energies of the peaks in the \ac{INS} cut at $0.25\K$. Explicitly one has the spectrum
\begin{align}
  \label{eq:best-spectrum}
  E_0    &\equiv 0.000 \meV \  (E), &
  E_{3'} &= 0.754 \meV \ (E), \nonumber \\
  E_1    &= 0.382 \meV \ (A_2), &
  E_4    &= 1.802 \meV \ (T_2), \nonumber \\
  E_2    &= 0.530 \meV \ (T_1), &
  E_{4'} &= 1.802 \meV \ (E), \nonumber \\
  E_3    &= 0.806 \meV \ (T_2),
\end{align}
where the irreducible representation in $T_d$ of each level is indicated. We note that
the $E_4$ and $E_{4'}$ levels are very close in energy, but not exactly equal. The model also accurately reproduces the wave vector and temperature dependence of the INS data as can be seen in Fig.~\subref{fig:neutron}{(c),(d),(f)}. Additional comparisons to magnetization and \ac{INS} data can be found in the Supplemental Material~\cite{supp}. Some of the features of these energy levels can be better understood by adopting global quantization axes and
defining global pseudo-spin operators $\tilde{\vec{S}}_i$. Using the notation of Ref.~[\onlinecite{yan-2013-living-edge}], the model in the global basis is parametrized by four anisotropic exchanges $J_1$, $J_2$, $J_3$ and $J_4$.  The best fit parameters of Eq.~(\ref{eq:best}) correspond to the values~\cite{supp}
\begin{align}
\label{eq:best-global}
  J_{1} &= +0.587 \meV, &
  J_{2} &= +0.573 \meV, \nonumber \\
  J_{3} &= -0.011 \meV, &
  J_{4} &= -0.117 \meV.
\end{align}
Since $J_1 \sim J_2 \equiv J$ and $J_3 \sim 0$ to a fair approximation, these fitted parameters describe a Heisenberg antiferromagnet supplemented with large (indirect) DM interaction $D \equiv \sqrt{2} J_4 \sim -0.28J$~\cite{canals-lacroix-2008-ising,supp} and negligible symmetric anisotropies. We can thus understand the $E$ doublet ground state as an extension of the pair of $S=0$ singlets that form the ground state in the Heisenberg limit~\cite{kimura-2014-breathing}. Similarly, the approximate quintet $E_4 \sim E_{4'}$ can be mapped to the high energy, five-fold degenerate $S=2$ states of the antiferromagnetic Heisenberg model. Indeed, when only Heisenberg and \ac{DM} interactions are present these remain exact eigenstates and degenerate, leaving only the small symmetric anisotropies to provide any splitting. While this mapping is appealing, there are key differences; for example, the three $S=1$ triplets present in the Heisenberg model are strongly mixed by the \ac{DM} interactions.

\shortsection{Discussion} The physics at very low temperatures, $T \ll E_1$, should be primarily controlled by the ground $E$ doublet. The states of this $E$ doublet,  $\ket{\pm}$,  are rather exotic. As in the Heisenberg limit, they are largely non-magnetic, carrying a uniform (scalar) spin-chirality $\kappa \equiv \avg{\tilde{\vec{S}}_i \cdot (\tilde{\vec{S}}_j \times \tilde{\vec{S}}_k)} \sim 0.4$ on each triangle of the tetrahedron~\cite{kimura-2014-breathing}. However, due to the large \ac{DM} interaction, the states additionally acquire \ac{AIAO} moments. This is generically expected as the \ac{AIAO} moments and the uniform spin-chirality transform identically under the tetrahedral symmetry~\cite{kotov-zhitomirsky-2004-weak-dimer, kotov-2004-patterns}. Explicitly, the projection of a pseudo-spin $\vec{S}_i$ in the local basis into the $E$ doublet takes the form $\bra{{\pm}}\vec{S}_i\ket{{\pm}} = \pm \lambda \vhat{z}$ with $\lambda \sim 0.13$ for the parameters of Eq.~(\ref{eq:best}) and $\bra{{\pm}} \vec{S}_i\ket{{\mp}}=0$. These \ac{AIAO} moments are \emph{octupolar} in character, with the net magnetic moment on each tetrahedron vanishing.  Due to the smallness of the inter-tetrahedron interactions we thus expect \ac{BYZO} to be an octupolar paramagnet at temperatures much smaller than $E_1$. Direct signatures of this unusual paramagnetic state may appear in more indirect magnetic probes, such as the non-linear susceptibilities.

Going to even lower temperatures one can potentially see indications of collective behavior of the small tetrahedra. Depending on the structure of the inter-tetrahedron interactions, a variety of states could be stabilized, such as weak \ac{AIAO} order or non-magnetic valence bond solid phases~\cite{kotov-zhitomirsky-2004-weak-dimer, kotov-2004-patterns}. Tantalizing hints of the onset of such correlations may already be present in the experimental data. We note that the \ac{INS} data is slightly broader than calculated instrumental resolution (by $\sim 0.01 \meV$) which may be suggestive of weak dispersion, while the specific heat data of~\citet{kimura-2014-breathing} shows a slight upturn below $\sim 500\mK$ that is not explained by the single-tetrahedron model. We thus suspect that the current lowest temperatures explored in \ac{BYZO} are at the threshold of observing such inter-tetrahedron correlations and possibly even ordering of these $E$ doublet states. Given the complete characterization of the intra-tetrahedron physics presented in this work, we feel the field is well poised to push the study of \ac{BYZO} to even lower temperatures and explore such inter-tetrahedra physics.

\begin{acknowledgements}
We thank J. Y. Y. Lin for the help with the data reduction.  A.D.C., M.D.L. and L.S.W. thank A. Chernyshev, P. Maksimov, G. Ehlers, and I. Zaliznyak for useful discussions. We thank K. Kimura and S. Nakatsuji for kindly providing their data from Ref.~[\onlinecite{kimura-2014-breathing}]. The research at the Spallation Neutron Source (ORNL) is supported by the Scientific User Facilities Division, Office of Basic Energy Sciences, U.S. Department of Energy (DOE). AFM was supported by the U. S. Department of Energy, Office of Science, Basic Energy Sciences, Materials Sciences and Engineering Division. Research supported in part by the Laboratory Directed Research and Development Program of Oak Ridge National Laboratory, managed by UT-Battelle, LLC, for the U. S. Department of Energy. The work at U. of Waterloo was supported by the NSERC of Canada, the Canada Research Chair program (M.J.P.G., Tier 1), the Canadian Foundation for Advanced Research and the Perimeter Institute (PI) for Theoretical Physics. Research at PI is supported by the Government of Canada through Industry Canada and by the Province of Ontario through the Ministry of Economic Development \& Innovation.

J.G.R. and L.S.W. contributed equally to this work.
\end{acknowledgements}

\clearpage

\addtolength{\oddsidemargin}{-0.75in}
\addtolength{\evensidemargin}{-0.75in}
\addtolength{\topmargin}{-0.725in}

\newcommand{\addpage}[1] {
    \begin{figure*}
      \includegraphics[width=8.5in,page=#1]{supp.pdf}
    \end{figure*}
}

\addpage{1}
\addpage{2}
\addpage{3}
\addpage{4}
\addpage{5}
\addpage{6}
\addpage{7}

\end{document}